\documentstyle[12pt]{article}

\begin{document}
\topmargin 0pt
\oddsidemargin 5mm

\setcounter{page}{0}

\begin{titlepage}

\begin{flushright}
OU-HET 348\\
TIT/HET-448\\
hep-th/0006025\\
June 2000
\end{flushright}
\bigskip

\begin{center}
{\LARGE\bf
Supersymmetric Nonlinear Sigma Models 
\footnotetext[1]{
Talk given at the International Symposium 
on ``Quantum Chromodynamics and Color Confinement" (Confinement 2000) 
, 7-10 March 2000, RCNP, Osaka, Japan.}

}\vspace{10mm}

\bigskip
{\renewcommand{\thefootnote}{\fnsymbol{footnote}}
{\large\bf Kiyoshi Higashijima$^a$\footnote{
     e-mail: higashij@phys.sci.osaka-u.ac.jp.}
 and Muneto Nitta$^b$\footnote{
e-mail: nitta@th.phys.titech.ac.jp}
}}

\setcounter{footnote}{0}
\bigskip

{\small \it
$^a$Department of Physics,
Graduate School of Science, Osaka University,\\
Toyonaka, Osaka 560-0043, Japan\\
$^b$Department of Physics, Tokyo Institute of Technology, 
Oh-okayama, \\ Meguro, Tokyo 152-8551, Japan\\
}
\end{center}
\bigskip

\begin{abstract}
Supersymmetric nonlinear sigma models 
are formulated as gauge theories. Auxiliary chiral superfields 
are introduced to impose supersymmetric constraints of F-type. 
Target manifolds defined by F-type constraints are always 
non-compact. In order to obtain nonlinear 
sigma models on compact manifolds, we have to introduce gauge 
symmetry to eliminate the degrees of freedom in non-compact 
directions. All supersymmetric nonlinear sigma models defined on 
the hermitian symmetric spaces are successfully formulated as 
gauge theories.
\end{abstract}

\end{titlepage}

\section{Introduction}
Two dimensional (2D) nonlinear sigma models and four dimensional 
non-abelian gauge theories have several similarities. 
Both of them enjoy the property of the asymptotic freedom. 
They are both massless in the perturbation theory, 
whereas they acquire the mass gap or the string tension in the 
non-perturbative treatment. 
Although it is difficult to solve QCD in analytical way, 
2D nonlinear sigma models can be solved by the large $N$ expansion 
and helps us to understand various non-perturbative phenomena 
in four dimensional gauge theories. 

In the nonperturbative treatment of nonlinear $\sigma$ models, 
auxiliary field method play an important role. As an example, 
let us consider the supersymmetric 2D nonlinear $\sigma$ model with 
$O(N)$ symmetry. The bosonic field $\vec{A}(x)$ takes 
the value on the real 
$N-1$ dimensional sphere $S^{N-1}$ with a radius ${\sqrt{N}\over g}$. 
The corresponding $N$ fermionic field $\vec{\psi}(x)$ is 
a Majorana (real) 
spinor which has the four-fermi type interaction with $O(N)$ symmetry 
and is called the Gross-Neveu model. 
The supersymmetric nonlinear sigma model 
with $O(N)$ symmetry is defined simply as a combination 
of these two models
\[
 {\cal L}(x)=
  {1\over 2}\{(\partial\vec{A})^2+
  \bar{\vec{\psi}}i\partial\llap /\vec{\psi}\}
  +{g^2\over 8N}(\bar{\vec{\psi}}\vec{\psi})^2 \label{eqn:onsigma}
\]
where $\vec{A}^2={N\over g^2}$ and $\vec{A}\cdot\vec{\psi}=0.$

This model enjoys three kinds of symmetry: $O(N)$ symmetry, 
discrete chiral symmetry $\psi \rightarrow \gamma_5\psi$ 
and the supersymmetry which mixes bosonic and fermionic fields. 
We can find the solution in the large $N$ limit where 
$N \rightarrow \infty$ with $g$ fixed. In order to study the phase 
structure, it is crucial to introduce auxiliary superfields 
defined in the superspace with the bosonic coordinate $x^{\mu}$ 
and the two component fermionic coordinate $\theta$
$$\Phi_0(x,\theta)=A_0(x)+\bar{\theta}\psi_0(x)
+{1\over 2}\bar{\theta}\theta F_0(x).$$ 
$A_0(x)\sim \vec{\bar{\psi}}\cdot\vec{\psi}$ describes the scalar 
bound state of two fermions as the auxiliary field of the 
Gross-Neveu model. $\psi_0(x)\sim \vec{A}\cdot\vec{\psi}$ describes 
the fermionic bound state of the fermion and the boson. 
$F_0(x)$ is the Lagrange multiplier for the constraint: 
$\vec{A}^2={N\over g^2}$ 
of the bosonic nonlinear $\sigma$ model. 
These auxiliary fields also play 
the role of order parameters of various symmetry. 
With the auxiliary field, the nonlinear lagrangian reduces to 
a very simple form
\[
 {\cal L}(x) =
   {1\over 2}\{(\partial\vec{A})^2
   + \bar{\vec{\psi}}i\partial\llap /\vec{\psi}+{\vec F}^2\}  
   + \left(2A_0{\vec A}\cdot{\vec F}-A_0\bar{\vec{\psi}}\vec{\psi}
  + F_0\vec{A}^2-\bar{\psi_0}{\vec\psi}\vec{A}
  - \bar{\vec\psi}\psi_0\vec{A}\right)-{N\over g^2}F_0.
\]
In the large $N$ field theory, we take into account the dominant 
vacuum fluctuations of $O(N)$ vector fields $\vec{A}, \vec{\psi}$ 
and $\vec{F}$ which can be integrated out easily since the lagrangian 
is simply of quadratic form in these variables.
As is shown in the table \ref{tbl:twod}, $O(N)$ symmetry recovers and 
$\vec{A(x)}$ acquire a mass proportional to $\langle A_0\rangle$. 
Chiral symmetry breaks down and fermions have the same mass 
with bosons.\\

\bigskip
\begin{table}
\caption{\bf Phase Structure of the $O(N)$ model}
\label{tbl:twod}
\begin{center}
\begin{tabular}{|c||c|c|c|}
\hline
symmetry & order parameter& perturbative & nonperturbative\\
\hline\hline
$O(N)$ & $\langle\vec{A}\rangle$ & $\times$ & $\bigcirc$\\
chiral symmetry & $\langle A_0\rangle$ &$\bigcirc$ & $\times$\\
supersymmetry &$\langle F_0\rangle$ &$\bigcirc$ &$\bigcirc$\\
\hline
\end{tabular}
\end{center}
\end{table}
In this talk, we generalize the auxiliary field formulation to 
nonlinear $\sigma$ models with ${\cal N}=2$ supersymmetry in 
two dimensions, which is equivalent to ${\cal N}=1$ supersymmetry 
in four dimensions.

\section{Nonlinear $\sigma$ Models in Four Dimensions}
When a global symmetry group $G$ breaks down to its subgroup $H$ 
by a vacuum expectation value $v=\langle \phi \rangle$, there appear 
massless Nambu-Goldstone (NG) bosons corresponding to broken generators 
in $G/H$. At low energies, interactions among these NG bosons are 
described by nonlinear $\sigma$ models~\cite{CCWZ}. 
In supersymmetric theories, target manifolds of nonlinear $\sigma$ models 
must be K\"{a}hler manifolds~\cite{Zu,WB}. 
A manifold whose metric is given by a K\"ahler potential $K(\bar{A},A)$
\begin{equation}
 g_{\bar{m}n} = {\partial^2K(\bar{A},A)\over 
   \partial\bar{A}^{\bar m}\partial A^n},\label{eqn:metric}
\end{equation}
ia called the K\"ahler manifold.
Since NG bosons must be scalar components of complex chiral superfields 
\begin{equation}
 \phi(x,\theta)=A(x)+\theta\psi(x)+{1\over 2}\theta^2F(x),
 \label{eqn:chiral}
\end{equation}
there are two possibilities. 
If the coset $G/H$ itself is a K\"ahler manifold, 
both ${\rm Re}\ A(x)$ and ${\rm Im}\ A(x)$ must be NG-bosons. 
The effective lagrangian in this 
case is uniquely determined by the metric of 
the coset manifold $G/H$~\cite{BKMU,IKK,BKY,Ku}. 
On the other hand, if the coset $G/H$ is 
a submanifold of a K\"ahler manifold, 
there is at least one chiral superfield whose real or 
imaginary part is not a NG-boson. 
This additional massless boson is called 
the quasi-Nambu-Goldstone(QNG) boson~\cite{BPY,KOY}. 
In this case, the K\"ahler metric in the direction of QNG boson is 
not determined by the metric of its subspace $G/H$, and the effective 
lagrangian is not unique.
In this article, we will confine ourselves 
to the case of K\"ahler $G/H$. 
The lagrangian of the nonlinear $\sigma$ model 
on a K\"ahler manifold is
\begin{eqnarray}
 {\cal L}(x) &=
   g_{\bar{m}n}\partial_{\mu}\bar{A}^{\bar m}\partial^{\mu}A^n
   +{i\over 2} g_{\bar{m}n}
    \left(\bar{\psi}^{\bar m}\sigma^{\mu}(D_{\mu}\psi)^n+
     \psi^n\bar{\sigma}^{\mu}(D_{\mu}\bar{\psi})^{\bar m}\right)
    \nonumber \\
   &+{1\over 4}R_{k\bar{m}l\bar{n}}(\psi^k\psi^l)(\bar{\psi}^{\bar m}
     \bar{\psi}^{\bar n}), \label{eqn:lagrangian}
\end{eqnarray}
where
\begin{equation}
 (D_{\mu}\psi)^n = (\delta^n_m\partial_{\mu}
   +\Gamma^n_{lm}\partial_{\mu}A^l)\psi^m, \qquad
 \Gamma^n_{lm} = g^{n\bar{k}}\partial_l g_{\bar{k}m}.
    \label{eqn:derivative}
\end{equation}
Once we know the K\"ahler potential $K(\bar{A},A)$, we can calculate the 
metric by (\ref{eqn:metric}), the connection by (\ref{eqn:derivative}) 
and the lagrangian by (\ref{eqn:lagrangian}).
This lagrangian is suitable for perturbative calculations. 
For the nonperturbative study, however, 
the auxiliary field formulation is 
hoped for. 

\section{Auxiliary Field Formulation}
Let us start from a known example, the ${\bf C}P^{N-1}$ model. 
Chiral superfield $\phi_i(x,\theta)\ (i=1,2, \cdots, N)$ belongs 
to a fundamental representation of $G=SU(N)$ 
which is the isometry of ${\bf C}P^{N-1}$. 
We introduce $U(1)$ gauge symmetry 
\begin{equation}
 \phi(x,\theta) \longrightarrow \phi'(x,\theta)
  = e^{i \Lambda(x,\theta)} \phi(x,\theta) \label{eqn:projective}
\end{equation}
to require that $\phi(x,\theta)$ and $\phi'(x,\theta)$ are 
physically indistinguishable. 
With a complex chiral superfield, 
$e^{i \Lambda(x,\theta)}$ is an arbitrary complex number. 
$U(1)$ gauge symmetry is thus complexified to $U(1)^{\bf C}$. 
The identification 
$\phi \sim \phi'$ defines the complex projective space ${\bf C}P^{N-1}$.
In order to impose local $U(1)$ gauge symmetry, 
we have to introduce a $U(1)$ gauge field 
$V(x, \theta, \bar{\theta})$ with 
the transformation property 
$e^V \longrightarrow e^V e^{-i\Lambda + i\Lambda^*}$.
Then the lagrangian with a local $U(1)$ gauge symmetry is given by
\begin{equation}
 {\cal L} = 
   \int d^2\theta d^2\bar{\theta} (e^V \vec{\phi}^*\cdot\vec{\phi}
    -cV) \label{eqn:cpnlag}
\end{equation}
where the last term $V$ is called the Fayet-Iliopoulos D-term.
In this model, real scalar superfield $V(x,\theta,\bar{\theta})$ 
is the auxiliary superfield. 
The K\"ahler potential $K(\phi, \phi^*)$ 
is obtained by eliminating $V$ 
by using the equation of motion for $V$
\begin{equation}
  {\cal L} = \int d^2\theta d^2\bar{\theta} K(\phi,\phi^*) 
  = c\int d^2\theta d^2\bar{\theta} 
     \log{({\vec{\phi}}^*\cdot\vec{\phi})}.
  \label{eqn:kahlerpot}
\end{equation}
This K\"ahler potential reduces to the standard Fubini-Study metric of 
${\bf C}P^{N-1}$ 
\begin{equation}
  K(\phi,\phi^*)=c\log{(1+\sum_{i=1}^{N-1}\phi_i^*\phi_i)}
   \label{eqn:fsmetric}
\end{equation}
by a choice of gauge fixing
\begin{equation}
  \phi_N(x,\theta)=1.\label{eqn:cpgauge}
\end{equation}
The lagrangian of the ${\bf C}P^{N-1}$ model is obtained by substituting 
the K\"ahler potential (\ref{eqn:fsmetric}) to eqns. 
(\ref{eqn:metric}), (\ref{eqn:derivative}) and (\ref{eqn:lagrangian}).
The global symmetry $G=SU(N)$, the isometry of the target space
${\bf C}P^{N-1}$, is linearly realized on our $\phi_i$ fields and our 
lagrangian (\ref{eqn:cpnlag}) with auxiliary field $V$ is manifestly 
invariant under $G$. The gauge fixing condition (\ref{eqn:cpgauge}) is
not invariant under $G=SU(N)$ and we have to perform an appropriate 
gauge transformation simultaneously to compensate the change of
$\phi_N$ caused by the $SU(N)$ transformation. Therefore the global 
symmetry $G=SU(N)$ is nonlinearly realized in the gauge fixed theory. 
In this sense, our lagrangian (\ref{eqn:cpnlag}) use the linear 
realization of $G$ in contrast to the nonlinear lagrangian in terms 
of the K\"ahler potential which use the nonlinear realization of $G$.

Similarly, we introduce $M\ (M<N)$ copies 
$\phi_j \ (j=1,2,\cdots,M)$ of fundamental 
representations of $SU(N)$ and impose the local gauge symmetry 
$U(M)$ to identify $\Phi U\sim \Phi\ (U\in U(M))$ where 
$\Phi=(\phi_1,\phi_2,\cdots,\phi_M)$ is an $N\times M$ matrix of 
chiral superfields. Then we obtain the nonlinear $\sigma$ model 
on the Grassmann manifold $SU(N)/SU(N-M)\times U(M)$. 
The lagrangian with an auxiliary $U(M)$ gauge field $V$ reads
\begin{equation}
{\cal L} = \int d^2\theta d^2\bar{\theta}\, 
     \left( {\rm tr} \left(\Phi^{\dagger}\Phi e^V\right) 
            - c\,{\rm tr} V \right)\label{eqn:grassmann}
\end{equation}
As in the ${\bf C}P^{N-1}$ model, the auxiliary field $V$ can be 
eliminated by the use of its equation motion. We fix the gauge by
choosing
\[
\Phi = \pmatrix {{\bf 1}_M \cr 
                  \varphi}  
\]
where $\varphi$ is an $(N-M)\times M$ matrix valued chiral superfield.
Then the K\"ahler potential reads
\begin{equation}
  K(\varphi, \varphi^{\dagger})=
  c\log\det({\bf 1}_M+\varphi^{\dagger}\varphi).
\end{equation}
Again, the global symmetry $G=SU(N)$ is linearly realized on our fields
$\Phi$ although the gauge fixed  fields $\varphi$ is no longer a linear 
realization.

\section{Nonlinear Sigma Models with F-term Constraints}
The superspace in the four dimensional space-time consists of 
four bosonic coordinates $x^{\mu}\quad (\mu=0,1,2,3)$ 
and a four components Majorana (real) spinor, 
which is equivalent to a complex Weyl spinor 
$\theta^{\alpha}\quad (\alpha=1,2)$ and 
its hermitian conjugate $\bar{\theta}$. 
The chiral superfield defined by 
(\ref{eqn:chiral}) depends only on $\theta$ but not on $\bar{\theta}$.
On the other hand, $\phi^{\dagger}$ depend on $\bar{\theta}$ 
but not on $\theta$. 
If we can make $G$ invariant combinations of chiral superfields, 
it is possible to introduce in the lagrangian another term 
called the F-term which can be written as an integral over $\theta$.

\medskip 
Let us try to impose an F-term constraint on ${\bf C}P^{N-1}$ model. 
The simplest constraint consistent with the $U(1)$ gauge symmetry 
(\ref{eqn:projective}) is the quadratic equation 
\begin{equation}
  \vec{\phi}\cdot\vec{\phi}=0. \label{eqn:quadra}
\end{equation}
With this constraint, the invariance group of the lagrangian is no 
longer $SU(N)$ symmetry but its subgroup $O(N)$. 
Any nonvanishing value is forbidden on the righthand side 
because of the $U(1)$ symmetry. 
Solution of this F-term constraint: 
$\vec{\phi}^{\,2}=\vec{x}^2+(y+iz)\cdot (y-iz)=0$ 
written in terms of $\vec{\phi} = (x_i, y, z)$ is given by 
\begin{equation}
  y-iz = - {x^2 \over y+iz}=-{x^2 \over \sqrt{2}} 
   \label{eqn:solution}
\end{equation}
where we have chosen a specific gauge $y+iz=\sqrt{2}$.
If we have imposed the constraint (\ref{eqn:quadra}) to $\vec{\phi}$ 
without introducing the gauge symmetry, 
the resulting target space would be 
noncompact K\"ahler manifold with a QNG boson. 
The QNG boson is now gauged away by the gauge symmetry 
(\ref{eqn:projective}) as a gauge degree of freedom 
and has disappeared from the physical spectrum.
On substitution of the solution (\ref{eqn:solution}) 
of the constraint to Eq.~(\ref{eqn:kahlerpot}), 
we obtain the K\"ahler potential of this model 
\begin{equation}
  K(x,x^*) = \log{\left( 1+\sum_{i=1}^{N-2}x_i^*x_i
  +{1\over 4}\sum_{i,j=1}^{N-2}x_i^{*2}x_j^2\right)},
  \label{eqn:quadratic}
\end{equation}
which is known as the K\"ahler potential of the quadratic surface 
\[
  Q^{N-2}({\bf C})={SO(N)\over SO(N-2)\times U(1)}.
\]
The lagrangian of the model with auxiliary fields is simply 
obtained from Eq~(\ref{eqn:cpnlag}) 
by imposing the F-term constraint (\ref{eqn:quadra}) 
with a lagrange multiplier field $\phi_0(x, \theta)$
\begin{equation}
  {\cal L} 
  = \int d^2\theta d^2\bar{\theta} 
   (e^V \vec{\phi}^*\cdot\vec{\phi}-V) 
  +\left( \int d^2\theta \phi_0\vec{\phi}\cdot\vec{\phi} 
   + {\rm h.c.} \right)
\end{equation}

\medskip
Let us impose an F-term constraint to the model 
on the Grassmann manifold $SU(2N)/SU(N)\times U(N)$. 
Our basic field is the $2N\times N$
matrix valued chiral superfield $\Phi$ which transforms linearly 
under the global symmetry $SU(2N)$. 
In order to identify $\Phi$ with $\Phi U$ with $U\in U(N)$, 
we introduce the $U(N)$ gauge field $V$.
The simplest constraint is again 
the quadratic constraint $\Phi^T\Phi=0$.
Since this constraint transforms as 
the symmetric second rank tensor under the gauge group $U(N)$, 
we introduce the chiral superfield $\Phi_0$ 
which also transforms as a symmetric second rank tensor 
under the gauge group $U(N)$. 
With this constraint, the global symmetry $SU(2N)$ 
reduces to its subgroup $SO(2N)$. 
Thus we obtain the auxiliary field formulation of 
the $SO(2N)/U(N)$ model
\begin{equation}
{\cal L} = \int d^4\theta\, 
     \left( {\rm tr} \left(\Phi^{\dagger}\Phi e^V\right) 
            - c\,{\rm tr} V \right) 
    + \left( \int d^2\theta \;{\rm tr} 
           \left(\Phi_0 \Phi^T\Phi)\right) 
            + {\rm h.c.} \right).\label{eqn:so2n}
\end{equation}

\medskip
If we insert the simpletic structure 
\begin{equation}
  J= \pmatrix {{\bf 0}&{\bf 1}_N \cr
       -{\bf 1}_N &{\bf 0} }. \label{eqn:J}
\end{equation}
between $\Phi^T$ and $\Phi$, the global symmetry reduces to the
simpletic group $Sp(N)$. 
Therefore, the $Sp(N)/U(N)$ model is defined by
\begin{equation}
 {\cal L} = \int d^4\theta\, 
     \left( {\rm tr} \left(\Phi^{\dagger}\Phi e^V\right) 
            - c\,{\rm tr} V \right) 
    + \left( \int d^2\theta \;{\rm tr} 
           \left(\Phi_0 \Phi^TJ\Phi)\right) 
            + {\rm h.c.} \right).
\end{equation}
In this case, the chiral auxiliary field $\Phi_0$ transforms as an
antisymmetric second rank tensor of the gauge group $U(N)$.

Similarly, we can formulate supersymmetric nonlinear sigma models 
on hermitian symmetric spaces shown in the table \ref{tbl:hss} 
by using auxiliary fields~\cite{HN}. 
It should be noted that the third order and the 
fourth order polynomials appear as the F-term constraints 
in the case of exceptional groups.

\begin{table}
\caption{\bf Hermitian Symmetric Spaces}\label{tbl:hss}
\begin{center}
\begin{tabular}{|c|c|c|}
 \noalign{\hrule height0.8pt}
  Type & $G/H$ \\
 \hline
 \noalign{\hrule height0.2pt}
  AIII$_1$ & ${\bf C}P^{N-1} = SU(N)/SU(N-1)\times U(1)$\\
  AIII$_2$ & $G_{N,M}({\bf C}) = U(N)/U(N-M)\times U(M)$ \\
  BDI      & $Q^{N-2}({\bf C}) = SO(N)/SO(N-2)\times U(1)$ \\
  CI       & $Sp(N)/U(N)$ \\
  DIII     & $SO(2N)/U(N)$ \\
  EIII     & $E_6/SO(10)\times U(1)$ \\
  EVII     & $E_7/E_6 \times U(1)$ \\  
 \noalign{\hrule height0.8pt}
 \end{tabular}
\end{center}
\begin{footnotesize}
First three manifolds, 
${\bf C}P^{N-1}$, $G_{N,M}({\bf C})$ and $Q^{N-2}({\bf C})$ 
are called a projective space, a Grassmann manifold 
and a quadratic surface, respectively. 
\end{footnotesize}
\end{table}
\section{Quantum Legendre Transform}
If we impose only the symmetry, the nonlinear lagrangian
may depend on arbitrary function. 
For example, consider the simplest ${\bf C}P^{N-1}$ model. 
The following lagrangian with 
an arbitrary function $f$ is allowed by the global as well as 
the local symmetry
\begin{equation}
{\cal L} = \int d^2\theta d^2\bar{\theta} 
(f(e^V \vec{\phi}^*\cdot\vec{\phi})
 -cV). \label{eqn:cpnlaga}
\end{equation}
We can prove this arbitrariness disappears and (\ref{eqn:cpnlaga}) 
reduces to the simplest lagrangian (\ref{eqn:cpnlag}) discussed 
previously~\cite{HN2}. 
Namely, we can prove that
\begin{eqnarray}
  \int[dV] \exp [i 
     \int d^4\theta ( f (e^V \phi^{\dagger}\phi) -cV)] 
 = \int[dV] \exp [i \int d^4\theta (e^V \phi^{\dagger}\phi -cV)]
  \nonumber
\end{eqnarray}
by using a remarkable property of the quantum Legendre transform 
in supersymmetric theories which is valid for any vector superfields
$\sigma(x,\theta,\bar{\theta})$, $\Phi(x,\theta,\bar{\theta})$
\begin{eqnarray}
 \int [d\sigma] \exp\left[i \int d^4x d^4\theta  
  \;(\sigma \Phi - W(\sigma)) \right] 
 =\exp\left[i \int d^4x d^4\theta  \;U(\Phi) \right], \label{eqn:qlt}
\end{eqnarray}
where $U(\Phi) = \hat \sigma (\Phi) \Phi - W(\hat {\sigma}(\Phi))$ is 
the Legendre transform of $W$ and $\hat\sigma$ is the stationary point:
\[
{\partial \over \partial \sigma} 
  (\sigma \Phi - W (\sigma)) |_{\sigma =\hat \sigma} 
 = \Phi - \partial W (\hat \sigma) = 0 .
\]

\section*{Acknowledgments}
The work of M.N. is supported in part by 
JSPS Research Fellowships.


\begin{thebibliography}{99}
\bibitem{CCWZ}
S.~Coleman, J.~Wess and B.~Zumino,
Phys. Rev. {\bf 177} (1969) 2239; \\
C.~G.~Callan, S.~Coleman, J.~Wess and B.~Zumino,
Phys. Rev. {\bf 177} (1969) 2247.

\bibitem{Zu}
B.~Zumino,
Phys. Lett. {\bf 87B} (1979) 203; 
L.~Alvarez-Gaum\'{e} and D.~Z.~Freedman, 
Comm. Math. Phys. {\bf 80} (1981) 443.

\bibitem{BKMU}
M.~Bando, T.~Kuramoto, T.~Maskawa and S.~Uehara,
Phys. Lett. {\bf 138B} (1984) 94;
Prog. Theor. Phys. {\bf 72} (1984) 313;
Prog. Theor. Phys. {\bf 72} (1984) 1207.

\bibitem{IKK}
K.~Itoh, T.~Kugo and H.~Kunitomo,
Nucl. Phys. {\bf B263} (1986) 295;
Prog. Theor. Phys. {\bf 75} (1986) 386.

\bibitem{BKY}
M.~Bando, T.~Kugo and K.~Yamawaki,
Phys. Rep. {\bf 164} (1988) 217.

\bibitem{Ku}
T.~Kugo, 
Soryuusiron Kenkyuu (Kyoto) {\bf 95} (1997) C56; 
``{\em Supersymmetric Nonlinear Realization}'', 
SCGT96 Proceedings (World Scientific, 1996), 
ed. by J.~Nishimura and K.~Yamawaki, 
available in http://ekenwww.phys.nagoya-u.ac.jp/Scgt/proc/.

\bibitem{BPY}
W.~Buchm\"{u}ller, R.~D.~Peccei and T.~Yanagida, 
Phys. Lett. {\bf 124B} (1983) 67; 
Nucl. Phys. {\bf B227} (1983) 503.  

\bibitem{KOY}
T.~Kugo, I.~Ojima and T.~Yanagida, 
Phys. Lett. {\bf 135B} (1984) 402.

\bibitem{HN} 
K.~Higashijima and M.~Nitta, 
Prog. Theor. Phys. {\bf 103} (2000) 635, hep-th/9911139.

\bibitem{HN2} 
K.~Higashijima and M.~Nitta, 
Prog. Theor. Phys. {\bf 103} (2000) 833, hep-th/9911225.

\bibitem{WB}
J.~Wess and J.~Bagger,
``{\em Supersymmetry and Supergravity}'', 
Princeton Univ. Press, Princeton (1992).

\end{thebibliography}
\end{document}